\documentclass{article}
\usepackage{spconf,amsmath,graphicx}
\usepackage{subcaption}
\usepackage{booktabs, multirow}

\usepackage{hyperref}
\usepackage{cleveref}
\crefname{section}{Sec.}{Sections}
\crefname{figure}{Fig.}{Figs.}
\crefname{table}{Tab.}{Tabs.}
\crefname{equation}{Eq.}{Eqs.}
\crefname{appsec}{Appendix}{Appendices}
\usepackage{enumitem}
\setlist{nosep, leftmargin=14pt}

\usepackage{mwe} 

\title{Classification of Luminal Subtypes in Full Mammogram Images Using Transfer Learning}
 \name{Adarsh Bhandary Panambur$^{\star \dagger}$ \qquad Prathmesh Madhu$^{\star}$ \qquad Andreas Maier$^{\star}$}
 \address{$^{\dagger}$ Siemens Healthineers, Technology Excellence, Erlangen, Germany\\
 $^{\star}$ Friedrich-Alexander-Universit\"at Erlangen-N\"urnberg, Pattern Recognition Lab, Erlangen, Germany  }
\begin{document}
\ninept
\maketitle
\begin{abstract}
Automatic identification of patients with luminal and non-luminal subtypes during a routine mammography screening can support clinicians in streamlining breast cancer therapy planning. Recent machine learning techniques have shown promising results in molecular subtype classification in mammography; however, they are highly dependent on pixel-level annotations, handcrafted, and radiomic features. In this work, we provide initial insights into the luminal subtype classification in full mammogram images trained using only image-level labels. Transfer learning is applied from a breast abnormality classification task, to finetune a ResNet-18-based luminal versus non-luminal subtype classification task. We present and compare our results on the publicly available CMMD dataset and show that our approach significantly outperforms the baseline classifier by achieving a mean AUC score of 0.6688 and a mean F1 score of 0.6693 on the test dataset. The improvement over baseline is statistically significant, with a $p$-value of $p$\textless 0.0001. 

\end{abstract}
\section{Background}
\label{sec:intro}

During the last few decades, breast cancer has rapidly become a global healthcare burden, with an estimated 2.26 million new incidences and 0.68 million deaths in 2020 \cite{globocan}. Early detection and diagnosis of abnormalities in the breast might result in a better prognosis and eventually provide clinicians with better treatment options. In this regard, mammography is the gold standard breast cancer screening technology for women. Breast Imaging Reporting and Data System (BIRADS) is a tool used by radiologists to standardize and manage breast mammography screening examinations \cite{birads}. Abnormal findings such as mass and calcification are often regarded as high-risk indications and can reveal signs of malignancy, depending on their size, growth, and distribution. Based on the BIRADS score, a biopsy is recommended, where a small part of the tissue in the region of abnormality is removed and further analyzed using histopathological analysis. The pathologists then analyze the tissue using microscopy to confirm whether the finding is benign or malignant and if malignant, then further classify the finding into various pathological subtypes. Upon confirmation of malignancy, immunohistochemistry analysis is performed to investigate the molecular subtypes for finding the genetic expression of the cancer cells. The four main molecular subtype categories are luminal A, luminal B, human epidermal growth factor receptor 2 (HER2)-enriched, and triple-negative \cite {horvath2021molecular, cho2016molecular, boisserie2013correlation}. Finally, the clinicians use the molecular subtypes to prepare a personalized cancer treatment plan. Even though magnetic resonance imaging includes more information about the molecular subtypes, recent studies have shown imaging feature correlations between mammography imaging and molecular subtypes \cite {horvath2021molecular, cho2016molecular, boisserie2013correlation}.  
Deep learning (DL)-based computer-aided diagnosis (CAD) solutions have been shown to achieve good performances in medical image analysis and, in many cases, have achieved on-par performance with clinicians \cite{nagendran2020artificial}. Most of the tasks involving DL methods in mammography analysis revolve around the classification, segmentation, and detection of abnormalities such as calcification, mass, and their benignancy and malignancy. However, there is a lack of research involving the analysis of molecular subtype features in full mammogram images using DL-based methods. A CAD system that can detect these molecular signatures directly from mammography can potentially streamline the clinicians' workflow, especially in countries where it is the only standalone cancer diagnosis tool. Furthermore, including additional information of these subtypes can result in designing a sophisticated multi-task learning-based CAD system. 

\subsection{Previous Works}

Earlier works involving subtype analysis in full mammogram images mainly use handcrafted and radiomics features, where annotations from expert radiologists are acquired. These features are used to train machine learning models for classification \cite {son2020prediction, ma2022predicting, wu2019prediction}. Son, Jinwoo, et al. use a total of 129 radiomic features from 365 patients and classify the synthetic mammograms obtained from 3D breast tomosynthesis into luminal, HER2, and triple-negative subtypes \cite{son2020prediction}. Ma, Mengwei, et al. use the clinical information and 50 imaging features based on abnormal findings seen in mammograms and ultrasound imaging from 600 patients, to develop and compare different machine learning models for the subtype classification \cite{ma2022predicting}. The BIRADS information from MRI and mammography is converted into 82 imaging features from 363 patients and is used to develop a Decision Tree (DT) model by Wu, Mingxiang et al. \cite{wu2019prediction}. They show the combination of features extracted using both imaging modalities can boost the performance of the subtype classification model. By using BIRADS features, they indicate the generalization ability of the model, by being modality- and vendor-independent \cite{wu2019prediction}.  

All the previous studies involving the subtype classification in mammography using deep learning methods are trained on the region of interest (ROI) from full mammogram images \cite {ueda2021training, muramatsu2022intrinsic, zhang2021predicting}. Ueda, Daiju, et al. proposed the first DL method to predict the hormone receptor expressions responsible for the molecular subtypes \cite{ueda2021training}. They use an ensemble of VGG, Inception Net, and ResNet models for training and validation of their method using 1448 ROIs from mammogram images \cite{ueda2021training}. Zhang, Tianyu et al. use an attention-mechanism-based ResNet-50 which uses a combination of mammogram and ultrasound ROIs as input to predict the four molecular subtypes. They also highlight the performance of luminal and non-luminal subtypes in their work \cite{zhang2021predicting}. Muramatsu, Chisako, et al. propose a contrastive pretraining on 385 ROI images, and k-nearest neighbor classifier on top of the extracted features for the classification of four molecular subtypes \cite{muramatsu2022intrinsic}. The datasets used in all the earlier works involve mammograms with high-quality pixel-level labels annotated by experienced radiologists and are not publicly available. 

\subsection{Research Goals}
In this work, we investigate the feasibility of classifying the full mammogram images with only image-level labels into luminal and non-luminal subtypes using a convolutional neural network (CNN), trained and validated using the publicly available, Chinese Mammography Dataset (CMMD) \cite{cui2021chinese}. The main contributions of our study are: 1. Analysis of the complexity in training a CNN model, for luminal and non-luminal subtype classification; 2. Selection of a suitable CNN model for transfer learning (TL) by comparing three different CNNs for abnormality classification tasks; 3. Selecting the multi-label, multi-class (MLMC) ResNet-18 (R18)-based model to classify the calcification, mass, benign and malignant findings for transfer learning; 4. Developing a luminal subtype classifier using the pre-trained model from the previous step. To the best of our knowledge, we show the first baseline results on the CMMD dataset for the luminal versus non-luminal subtype classification. 


\section{Methods}
\label{sec:methods}
\subsection{Data}

As mentioned previously, we use the publicly available CMMD dataset from the ``The Cancer Imaging Archive" (TCIA) \cite{cui2021chinese, clark2013cancer}. The dataset consists of 3,744 mammograms belonging to 1,775 patients from China. Each mammogram is labeled with the age of the patient, craniocaudal (CC)/mediolateral oblique (MLO) views, right/left breast, calcification/mass anomalies, and benign/malignant lesions. A subset of these mammograms from 749 patients with malignant lesions is labeled with luminal A, luminal B, HER2-enriched, and triple-negative subtypes, respectively. In this work, we aim to perform a luminal versus non-luminal classification study. Therefore, luminal A and luminal B subtypes are combined to form the `luminal' class. HER2-enriched and triple-negative are merged into the `non-luminal' class. We use the entire dataset for the abnormality classification tasks and the subset for the luminal versus non-luminal classification task.

\begin{table}[!ht]
    \centering
    \caption{Class sample distribution in the CMMD dataset. The entire dataset is split into approximately 90:10 for training and testing. 90\% of the dataset is used for the 5-fold cross-validation (CV) splits, where almost 80\% is used for training and 20\% is used for validation.}
    \label{tab:dataset}
    \begin{subtable}{\linewidth}
    \centering
        \begin{tabular}{c  c  c } 
         \toprule
                       & 5-fold CV Dataset          &  Test Dataset \\ 
        \midrule
         Luminal         & 956                        & 100            \\ 
        Non-Luminal         & 404                        & 38            \\ 
        \midrule
       Total           &  1360                      & 138         \\
        
         \bottomrule
        \end{tabular}
        \caption{Dataset used for luminal versus non-luminal classification task.}
        \label{tab:dataseta}
    \end{subtable}

    \begin{subtable}{\linewidth}
        \centering
        \begin{tabular}{c  c  c } 
    
        \toprule
                & 5-fold CV Dataset        &  Test Dataset \\  [0.5ex] 
         \midrule
         Benign          &  1004                    & 108           \\

        Malignant       &  2386                    & 246           \\ 
        
        \midrule

        Calcification   &  476                     &  48                    \\

        Mass            &  2076                    & 222                 \\

        Calc+Mass       &  838                     & 84                 \\

        \midrule
        Total           &  3390                    & 354                \\

         \bottomrule
        
        \end{tabular}
        \caption{Dataset used for the abnormality classification tasks.}
        
        \label{tab:datasetb}
    \end{subtable}
\end{table}

\cref{tab:dataset} shows the distribution of the class samples in both the datasets used in this study. We perform a 5-fold cross-validation (CV) in all experiments conducted in this study and test individual folds using the independent test dataset. 90 \% of the total dataset is used for CV and 10 \% is used as the independent test dataset. Approximately, 80\% of the CV dataset is used for training, and 20 \% is used for validation across all the folds. We ensure that all the folds consist of an almost equal distribution of the class samples and that there is no leakage of data samples belonging to the same patient between the training and validation sets. The mammograms in the CMMD dataset are converted from Digital Imaging and Communications in Medicine (DICOM) files into JPEG format. The quality of the saved images is preserved by scaling the pixel values using the window-width and window-level from the DICOM metadata. Furthermore, the black pixels outside the breast region are cropped-out to obtain only the breast view as the input to the CNN. The images are resized to 1326 x 512 using the average aspect ratio of all cropped images in the dataset to meet the computational requirements for model training.

\begin{figure*}[htb]

\begin{minipage}[b]{1.0\linewidth}
  \centering
  \centerline{\includegraphics[width=11.7cm, height=6.7cm]{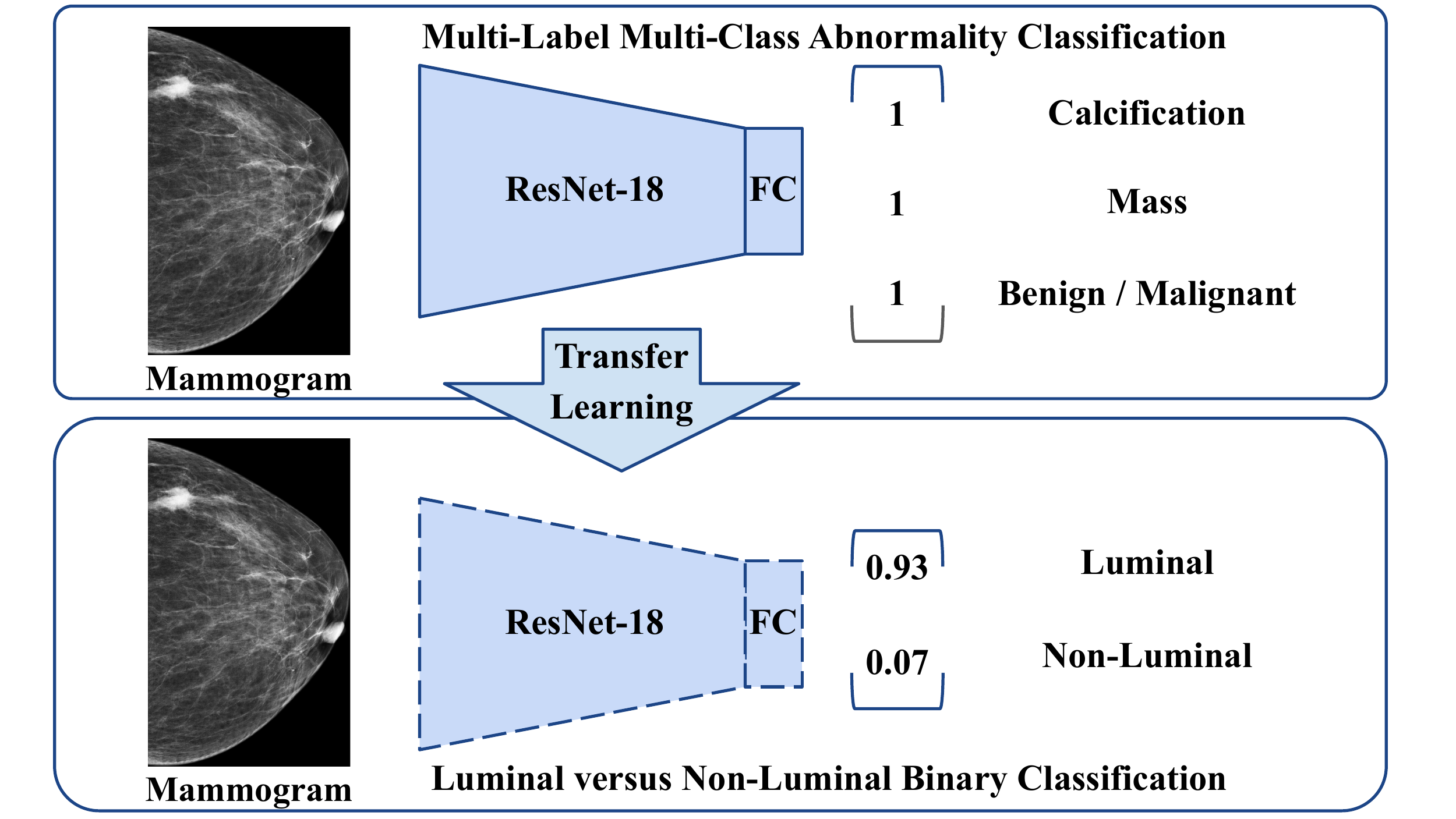}}

\end{minipage}
\caption{Overview of the transfer learning method: MLMC classification network is used for supervised pretraining for the mass, calcification, benign, and malignant classes. The pretrained network is then used for finetuning the CNN for the luminal subtype classification task.}
\label{fig1}
\end{figure*}

\subsection{Experimental Setup}

R18 pretrained on ImageNet is used as the standard CNN architecture across all the experiments conducted in this research. The CNN is selected after searching for the right balance between the number of training parameters and FLOPs to ensure faster training performance on large-resolution input images. During training, we perform four data augmentations, namely random horizontal flip (p=0.5), AugMix (p=0.2) \cite{hendrycks2019augmix}, random histogram equalization (p=0.4), and random erasing (p=0.1). We employ an early stopping criterion with a patience value of 10, to prevent the models from overfitting and achieve faster training times. All models are saved on the lowest validation loss to avoid using the overfitted models during testing. The experimental setup in this study is designed using the PyTorch framework.

\subsubsection{Baseline: Luminal Versus Non-Luminal Classification Task}

\label {sub:baseline}

At first, we perform the luminal versus non-luminal subtype classification, by refactoring the fully connected (FC) layer of R18 with two output nodes, representing the two classes. As the training is extremely unstable, we perform experiments by adding the dropout layer with various probabilities, eventually selecting a value of 0.3. A weighted random sampler is used during training to address the data imbalance problem in the subtype dataset. Based on our initial experiments, we see a significant gain in performance while using this sampling strategy. A batch size of 16 is used for all the experiments and a cross-entropy (CE) loss is optimized using the standard Adam optimization strategy with a learning rate of $1e^ {-5} $ and a weight decay of $5e^ {-3} $. It is seen in our experiments that using low learning rates can stabilize the training to some extent for this task.

\subsubsection {Abnormality Classification Tasks: Binary-Class And Multi-Label Multi-Class Task}

As the performance of CNN is poor on the test dataset and training instabilities are observed during the 5-fold CV on the first task, we investigate whether TL from a related task can boost the performance of the luminal classifier. For this purpose, we investigate three classifiers, namely, mass versus calcification, benign versus malignant, and a combined MLMC classifier that classifies mass, calcification, and its benignancy and malignancy. The goal here is to perform supervised training on the aforementioned tasks and select the right model to perform TL. As the imaging features correlated to the subtypes arise due to abnormalities such as mass and calcification \cite {horvath2021molecular, cho2016molecular, boisserie2013correlation}, we initially employ a mass versus calcification classifier. As many mammograms consist of both mass and calcifications, a multi-label binary classifier is used with a binary CE loss function. The benign versus malignant classification is then explored, as the molecular subtypes are the hormone receptor expressions arising from the malignant lesions. We employ a binary output on top of the FC layer and use a CE loss function. Finally, the MLMC classifier is employed with the intuition that a single network should be able to learn robust representations of both the mass versus calcification and benign versus malignant task. In this case, the FC layer is refactored to output three neurons, each representing the probability of one of the three classes. As only a subset of the mammograms with malignant lesions is labeled with the molecular subtypes, we do not include the subtype label in the MLMC task. A binary CE loss is employed again in this case. Based on extensive experimentation as reported in \cref{sec:results}, we propose to use the MLMC classifier for supervised pretraining. A learning rate of $1e^ {-4} $ and weight decay of $5e^ {-3} $ is used for the Adam optimizer.  

\subsubsection {Transfer Learning Task: Luminal Versus Non-Luminal Classification Task} 

The overview of the approach of the TL task can be seen in \cref{fig1}. We replace the fully connected layer from three to two, to match the luminal and non-luminal subtype classes. Each fold is initialized by the pre-trained weights from each fold in the MLMC task, and all layers in the R18 model are finetuned to ensure optimal performance. The remaining settings from \cref{sub:baseline} are retained for a fair comparison between the two methods.

\section{Results and Discussion}
\label{sec:results}

\begin{figure*}[htb]

\begin{minipage}[b]{1.0\linewidth}
  \centering
  \centerline{\includegraphics[width=13.2cm, height=6cm]{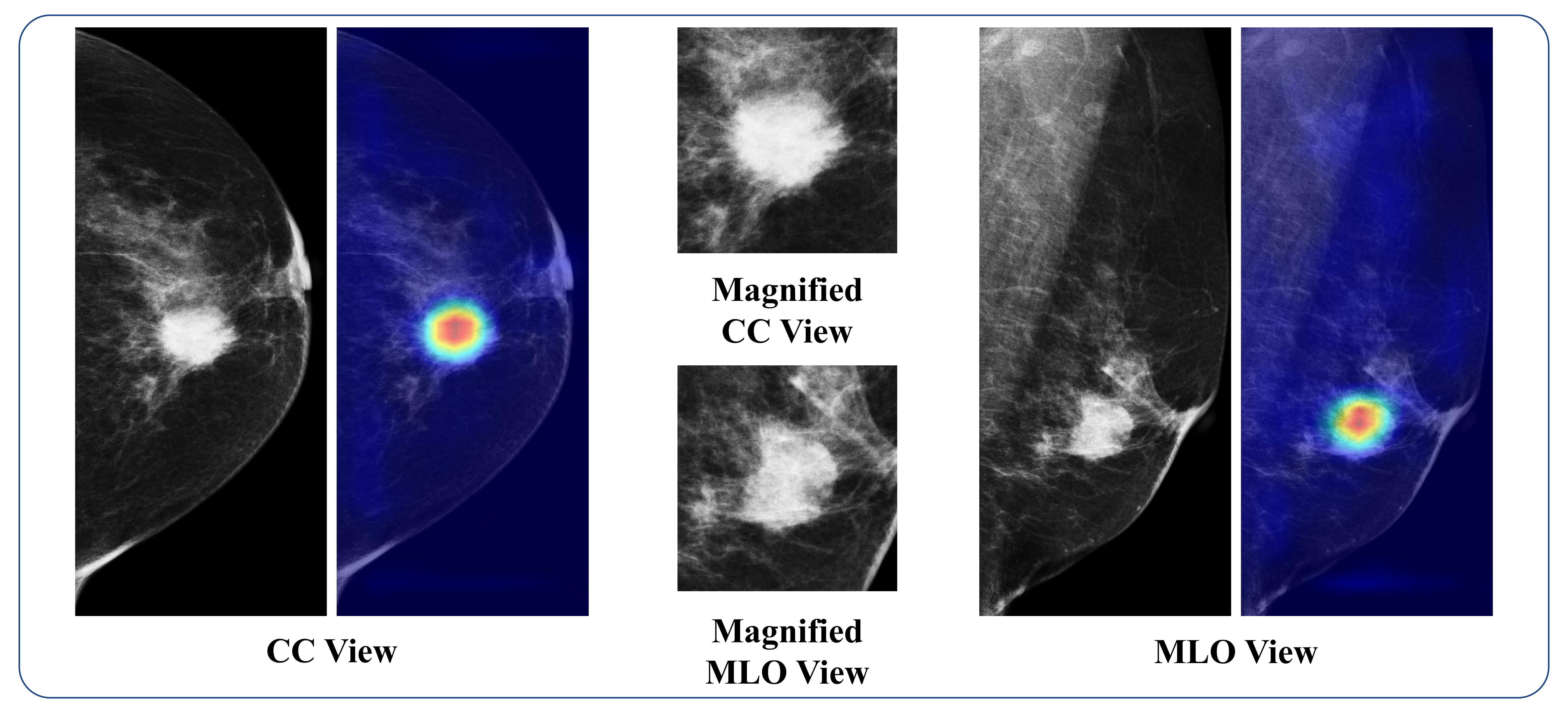}}
\end{minipage}
\caption{Grad-CAM visualizations and magnified ROI of the CC and MLO view from the mammogram of a 58-year-old breast cancer patient with both calcification and mass findings. The malignant lesion belongs to the luminal subtype.}
\label{fig:grad}
\end{figure*}
The per-class F1 score, total F1 score, and Area Under the Curve (AUC) score are used as the evaluation metrics for the luminal subtype classification. As there are three separate tasks in abnormality classification, only the per-class F1 score is used to evaluate the performance of the models. The model from each of the five folds is used for testing and the mean and standard deviation values are reported. 

From \cref{tab:restable2}, it is clear that the luminal versus non-luminal classification model fails to learn good representations of the molecular subtypes from the full mammogram images. It achieves a mean F1 score of 0.4987 and a mean AUC of 0.5358, with a mean F1 score as low as 0.2099 for the non-luminal class. One potential reason for the inferior performance can be attributed to the class imbalance, especially with the high F1 score of the dominant luminal subtype class, even after using weighted random sampling. Moreover, there is no additional clinical information on the local appearance of calcification and mass in the CMMD dataset. For example, the presence of microcalcification clusters, mass with well-circumscribed or poorly-circumscribed margins, and spiculations are some of the imaging features which highlight the specific molecular subtype signatures \cite{cho2016molecular}. As this combination of clinical knowledge in the form of class label information is missing, the network fails to learn the correlations between the abnormalities and molecular subtypes. This can also be supported by the increased performance of the CNNs as reported in the literature when trained on smaller ROIs~\cite{ueda2021training, zhang2021predicting}.  

\begin{table}[t]
    \centering
    \caption{Classification performance on the test dataset. The mean values on the test dataset from 5-fold CV are reported. Standard deviations are reported in brackets. The best results are shown in bold.}
    \begin{subtable}{\linewidth}
        \centering
        \begin{tabular}{c  c  c } 
         \toprule
        Class                & Binary-class F1         & MLMC F1 \\ 
        \midrule
         Calcification       & 0.7765 (0.0116)                   & \textbf{0.8025     (0.0190)}           \\ 
        
        Mass                 & 0.9233  (0.0036)                   & \textbf{0.9238   (0.0039)}    \\         
        
        Benign               & 0.4079  (0.083)                  &  \textbf{0.5061    (0.0860)}                 \\
        
        Malignant           & 0.8313  (0.0039)                   &  \textbf{0.8480   (0.0079)}          \\          
         \bottomrule
        \end{tabular}
        \caption{Binary versus MLMC results.}
        \label{tab:binmlmc}
    \end{subtable}\\%
    \begin{subtable}{\linewidth}
        \centering
        \begin{tabular}{c  c  c } 
    
        \toprule
                                & Baseline               &  Transfer learning \\ [0.5ex] 
         \midrule
        Luminal F1              & 0.7864 (0.0667)      & \textbf{0.8188     (0.0245)}       \\ 
        
        Non Luminal F1          & 0.2099 (0.1749)     & \textbf{0.5199 (0.0253)}       \\

        F1 Score                & 0.4987 (0.0647)      &  \textbf{0.6693  (0.0184)}       \\
        
        AUC                     & 0.5358 (0.0300)      & \textbf{0.6688     (0.0210)}    \\
         \bottomrule
        
        \end{tabular}
        \caption{Baseline versus Transfer learning results.}
        \label{tab:restable2}
    \end{subtable}
\end{table}




In comparison, it is easier for the network to classify the calcification, mass, benign, and malignant features. This can be seen quantitatively in \cref{tab:binmlmc} and might be due to the global appearance of calcification and mass on mammograms. Calcifications appear as bright, high-intensity, white spots on mammograms, while the global morphological features of most of the mass abnormalities are distinct. Therefore, we can hypothesize that these abnormalities are representative of their respective class labels. The same applies to malignant and benign classes as well. Moreover, we observe the network works better when we train a single network to classify all the abnormalities. Although the performance of the mass classification stays the same, we obtain more than 2\%, 9\%, and 1\% performance boost for the calcification, benign and malignant classes, respectively. We find these gains in performance to be statistically significant while using a t-test, with a two-tailed $p$-value of $p$$<$0.0001. Hence, we use the MLMC model for finetuning the luminal classifier.  

We achieve a mean AUC of 0.6688 and a mean F1-score of 0.6693 in the TL task. A high-performance boost with statistical significance ($p$$<$0.0001) in comparison to the baseline is obtained by performing TL, and this trend is seen across all the 5-folds. This can be observed in \cref{tab:restable2}, where we see that the increase in the overall performance can be attributed to the gains in the non-luminal class, showing the network no longer favors the majority class. The clinical information in the form of missing additional labels in the baseline models is replaced by rich representations of the same information learned from the supervised pretraining in the MLMC task. 
We further analyze the Gradient-Weighted Class Activation Mappings (Grad-CAM) to visualize the attention of the MLMC network. \cref{fig:grad} shows an example of the Grad-CAM obtained from the MLMC model. The suspected ROI is comparably found by the Grad-CAM in both CC and MLO views. Even though the network predictions are the same as the input class label, and the ROI looks visually suspicious by clinical definitions of mass and calcification, it is difficult to use these as weak localization maps for further analysis due to the absence of lesion location data. 

%
%
%
\section{Conclusion and Future Work}
\label{sec:typestyle}
In this work, we investigate the luminal versus non-luminal subtype classification on full mammogram images using only image-level labels. We provide initial insights into the issues with CNNs for this task and also show that TL from a robust MLMC abnormality classification model can significantly boost the performance of the subtype classification. We achieve a mean AUC and F1 score of 0.6688 and 0.6693 using TL in comparison to 0.5358 and 0.4987 while not using TL, respectively. These initial results on the luminal classification tasks are promising and reiterate the need for more active research in this sub-area of breast cancer to eventually develop a standalone multi-task CNN model for analysis of all types of breast abnormalities. In the future, we plan to validate the Grad-CAM localization results from a subset of this data with the help of a clinician and use it as the test dataset. We then aim to use the Grad-CAM generated localization maps to develop a weakly-supervised ROI classifier model and also extract radiomics features for further comparison with the current methods in luminal subtype analysis.

\section{Compliance with ethical standards}
\label{sec:ethics}

This research study is conducted using the open access data available in ``The Cancer Imaging Archive" (TCIA) \cite{cui2021chinese, clark2013cancer} and consists of human subject data. 
The license associated with the open-access data confirms the non-requirement of ethical approval.

\bibliographystyle{IEEEbib}
\bibliography{refs}

\end{document}